# The high-pressure phase of boron, γ-$B_{28}$: disputes and conclusions of 5 years after discovery.

Artem R. Oganov[1,2,*], Vladimir L. Solozhenko[3], Carlo Gatti[4], Oleksandr O. Kurakevych[3], Yann Le Godec[5]

*Corresponding author

[1] *Department of Geosciences, Department of Physics and Astronomy, and New York Center for Computational Science, Stony Brook University, Stony Brook, NY 11794-2100, USA.*

[2] *Geology Department, Moscow State University, 119992 Moscow, Russia.*

[3] *LSPM-CNRS, Université Paris Nord, 93430 Villetaneuse, France*

[4] *CNR-ISTM Istituto di Scienze e Tecnologie Molecolari, via Golgi 19, 20133 Milano, Italy.*

[5] *IMPMC, Université P&M Curie, 75005 Paris, France*

**Abstract**

**γ-$B_{28}$ is a recently established high-pressure phase of boron. Its structure consists of icosahedral $B_{12}$ clusters and $B_2$ dumbbells in a NaCl-type arrangement $(B_2)^{δ+}(B_{12})^{δ-}$ and displays a significant charge transfer δ~0.5-0.6. The discovery of this phase proved essential for the understanding and construction of the phase diagram of boron. γ-$B_{28}$ was first experimentally obtained as a pure boron allotrope in early 2004 and its structure was discovered in 2006. This paper reviews recent results and in particular deals with the contentious issues related to the equation of state, hardness, putative isostructural phase transformation at ~40 GPa, and debates on the nature of chemical bonding in this phase. Our analysis confirms that (a) calculations based on density functional theory give an accurate description of its equation of state, (b) the reported isostructural phase transformation in γ-$B_{28}$ is an artifact rather than a fact, (c) the best estimate of hardness of this phase is 50 GPa, (d) chemical bonding in this phase has a significant degree of ionicity. Apart from presenting an overview of previous results within a consistent view grounded in experiment, thermodynamics and quantum mechanics, we present new results on Bader charges in γ-$B_{28}$ using different levels of quantum-mechanical theory (GGA, exact exchange, and HSE06 hybrid functional), and show that the earlier conclusion about significant degree of partial ionicity in this phase is very robust. Additional insight into the nature of partial ionicity is obtained from a number of boron structures theoretically constructed in this work.**

## I. Introduction.

Boron is an element with interesting and complex chemistry, many details of which are still not well understood. At pressures below 89 GPa it adopts structures based on icosahedral $B_{12}$ clusters with multicenter bonds within the icosahedra and 2-centre and 3-centre bonds between the icosahedra. At least 16 crystalline allotropes have been reported [1], but crystal structures were determined only for 4 modifications and most of the reported phases are likely to be boron-rich borides rather than pure elemental boron [1-3]. Until 2007, it was the last light element, for which the ground state was not known even at ambient conditions (the debate whether α-$B_{12}$ or β-$B_{106}$ is stable at ambient conditions, was finally resolved in 2007-2009 by *ab initio* calculations of three different groups [4-6], which all concluded in favor of β-$B_{106}$ and against common intuition that favored the much simpler α-$B_{12}$ structure). Among the reported phases of boron, probably only four correspond to the pure element [1,2]: rhombohedral α-$B_{12}$ and β-$B_{106}$ phases (with 12 and 106 atoms in the unit cell, respectively), tetragonal T-192 (with 190-192 atoms/cell), and the newly discovered γ-$B_{28}$ – see Fig. 1.

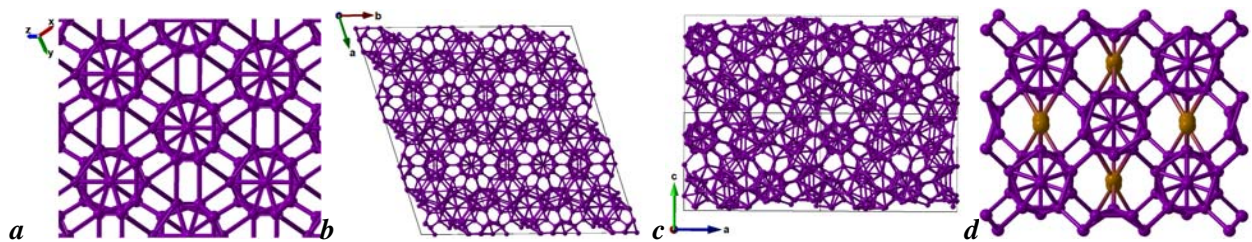

**Fig.1.** Crystal structures of boron polymorphs. (*a*) α-$B_{12}$, (*b*) β-$B_{106}$, (*c*) T-192, (*d*) γ-$B_{28}$. Reproduced from [7,8] with modifications.

The history of discovery of γ-$B_{28}$ has a twist. In 1964 R.H. Wentorf found that both β-$B_{106}$ and amorphous boron transformed into some other phase at pressures above 10 GPa and temperatures 1800-2300 K [9]. Wentorf reported a qualitative diffraction pattern of the product, but could not determine the structure or even the unit cell parameters, and, perhaps most seriously (in view of boron's sensitivity to impurities – e.g., such compounds as $PuB_{100}$ or $YB_{66}$ are known), he did not determine the chemical composition of his material. For that time, it was a state-of-the-art work, but nevertheless it was not accepted by the community, and Wentorf's paper remained essentially uncited for over 40 years and his diffraction data were deleted from the Powder Diffraction File Database. However, now it can be stated that with good likelihood Wentorf had synthesized the phase now known as γ-$B_{28}$ [7], in mixture with other phase(s). Another major result came from J. Chen and V.L. Solozhenko, who independently (in February and April 2004, respectively) found a new phase of pure boron at pressures above 10-12 GPa and temperatures above 1500 K. Although Chen managed to determine the unit cell parameters of the new phase (orthorhombic cell with a=5.0544 Å, b=5.6199 Å, c=6.9873 Å), neither he nor Solozhenko succeeded in solving its structure. In 2006, Chen posed this problem to Oganov, with the idea that Oganov's USPEX method for predicting crystal structures [10] could be used for solving this problem. The structure (Fig. 1d) was solved within 1 day[1].

USPEX [10,11] is an *ab initio* evolutionary algorithm, which searches for the structure with the lowest theoretical thermodynamic potential and requires no experimental information. However, the use of experimental cell parameters as constraints simplifies search and we took advantage of it. From densities of other boron phases, we estimated the number of atoms in the cell to be between 24 and 32. Since this number has to be even to produce an insulating state, we considered cases of 24, 26, 28, 30 and 32 atoms/cell. Fig. 2 illustrates this search process by the sequence of lowest-energy structures in each generation for the 24-atom system. The first (random) generation did not contain any icosahedral structures. Increasingly large fragments of the icosahedra appear during calculation, until at the 11[th] generation the lowest-energy structure is found. Fig. 3 shows the lowest-energy structures for each number of atoms in the unit cell. The 28-atom *Pnnm* structure (Fig. 1d) was found even faster (in 4 generations) and had the lowest energy per atom, correct orthorhombic symmetry, and relaxed cell parameters and diffraction pattern in good agreement with experiment (Fig. 4).

---

[1] But it took much longer to publish the results. The paper was submitted to *Nature* on 27 January 2007 and it took 2 years to publish it (the paper came out on 28 January 2009). During this period, Oganov and Chen learned about Solozhenko's independent work and the two teams merged.

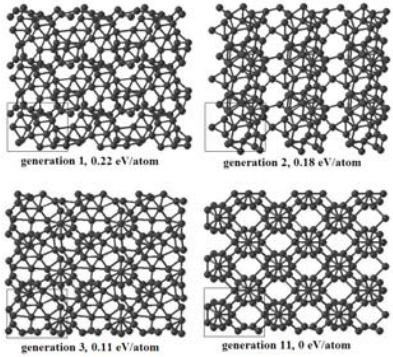

**Fig. 2.** Example of an evolutionary simulation (24-atom system at fixed cell parameters). Best structure at each generation is shown (with total energies relative to the final energy). From [7].

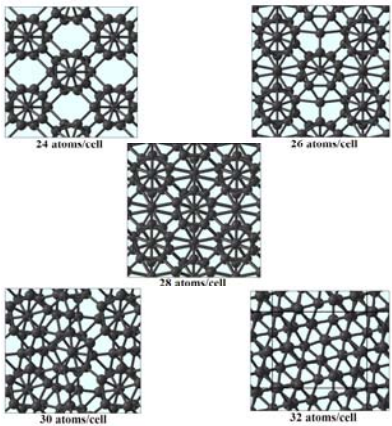

**Fig. 3.** Best structures with 24, 26, 28, 30, 32 atoms/cell [7]. Structures with 24 and 28 atoms/cell correspond to α-$B_{12}$ and γ-$B_{28}$, respectively, and contain full $B_{12}$ icosahedra.

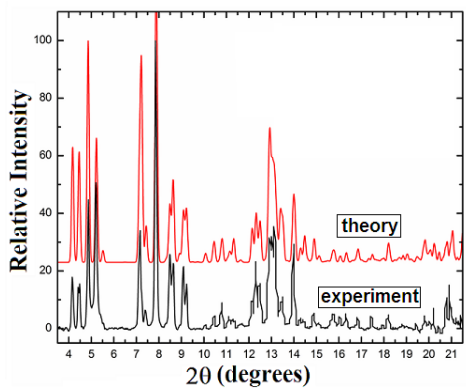

**Fig. 4.** Comparison of theoretical and experimental X-ray powder diffraction patterns of γ-$B_{28}$ (wavelength λ = 0.31851 Å) [7].

Table 1 gives predicted (at the DFT-GGA level of theory) structural parameters of γ-$B_{28}$ and two other stable boron phases with relatively simple structures (theoretical data on other phases are available from the first author). Excellent agreement with available experimental data can be seen.

Boron is the second hardest element after carbon (diamond, lonsdaleite, and M-carbon [10,12] allotropes), as was noted already in 1911 by Weintraub [13]. Furthermore, boron has one of the highest number densities (i.e. the number of atoms per unit volume) among all known substances at ambient conditions [14] – the record here also belongs to diamond (although there are many denser interesting structures of carbon indicated by theory [15]). γ-$B_{28}$ is the densest, and hardest, of all known boron phases. The best estimates of the hardness of β-$B_{106}$ and α-$B_{12}$ are 45 GPa [16] and 42 GPa [17], respectively, whereas for γ-$B_{28}$ the measured Vickers hardness is 50 GPa [18], which puts it among half a dozen hardest materials known to date. The high density of this phase is due to the close packing of the $B_{12}$ icosahedra - as in α-$B_{12}$, but with the "empty" space filled by the $B_2$ pairs.

While the paper [7] is the earliest publication by submission date (January 2007), paper [18] is the earliest in terms of publication date (November 2008). What these papers reported for the first time was the phase diagram of boron and structure, band gap, phonon dispersion curves, infrared spectrum, dielectric constants and peculiar chemical bonding of γ-$B_{28}$ [7], as well as its superhardness [18]. Fig. 5 shows the phase diagram reported in [7] and confirmed by subsequent experiments (V.L. Solozhenko, unpublished results; J. Qin et al., http://arxiv.org/abs/1109.1115). Now we turn to discuss subsequent works.

**Table 1. Structures of stable boron phases (optimized at 1 atm), with Bader charges (Q).** Experimental data are in parentheses (Refs.7,19). From [7] (except that updated, slightly different, values for the GGA Bader charges are given for γ-$B_{28}$). $Q_{GGA}$ and $Q_{IAM}$ are Bader charges obtained self-consistently and using superposition of spherically averaged atomic densities, respectively.

| Wyckoff position | x | y | z | $Q_{GGA}$ | $Q_{IAM}$ |
|---|---|---|---|---|---|
| γ-$B_{28}$. Space group *Pnnm*. $a$=5.043 (5.054) Å, $b$=5.612 (5.620) Å, $c$=6.921 (6.987) Å | | | | | |
| B1 (4g) | 0.1702 | 0.5206 | 0 | +0.26 | +0.0250 |
| B2 (8h) | 0.1606 | 0.2810 | 0.3743 | -0.18 | -0.0153 |
| B3 (8h) | 0.3472 | 0.0924 | 0.2093 | +0.00 | +0.0035 |
| B4 (4g) | 0.3520 | 0.2711 | 0 | +0.07 | -0.0003 |
| B5 (4g) | 0.1644 | 0.0080 | 0 | +0.04 | -0.0011 |
| α-$B_{12}$. Space group $R\bar{3}m$. $a$=$b$=$c$=5.051 (5.064) Å, $α$=$β$=$γ$=58.04 (58.10)°. | | | | | |
| B1 (18h) | 0.0103 (0.0102) | 0.0103 (0.0102) | 0.6540 (0.6536) | +0.0565 | -0.0030 |
| B2 (18h) | 0.2211 (0.2212) | 0.2211 (0.2212) | 0.6305 (0.6306) | -0.0565 | +0.0030 |
| α-Ga structure. Space group *Cmca*. a=2.939 Å, b=5.330 Å, c=3.260 Å. | | | | | |
| B1 (8f) | 0 | 0.1558 | 0.0899 | 0 | 0 |

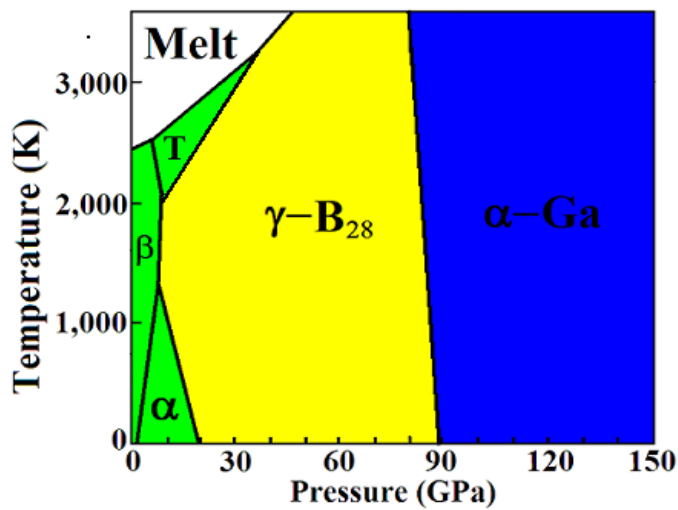

**Fig. 5.** Phase diagram of boron. Reproduced from [7].

Surprised by his findings, in December 2006 the first author of this paper sent the manuscript of [7] to a colleague, who as part of a team [20,21] confirmed the structure and superhardness of γ-$B_{28}$ two years later. The most original contribution of Zarechnaya *et al.* [20,21] is that they were able to grow micron-sized single crystals of this phase, but (as is often the case with boron research) encountered difficulties. Unfortunately, conditions of synthesis in [20,21] were suboptimal (e.g., metal capsules reacted with boron sample), and both papers contained ambiguities (see [22]). For instance, the estimated density differences between boron allotropes were wrong by an order of magnitude (Zarechnaya et al. [21] claimed that γ-$B_{28}$ is 1% denser than all other forms of boron, while it is actually 8.3% denser than β-$B_{106}$). Equally surprising is the statement that "*only $B_{28}$ contains additional B atoms in an intericosahedral space*" (it is well known β-boron [23,4-6] and the tetragonal phase T-192 [24] contain a very large number of intericosahedral atoms). These [20,21] and later papers by Dubrovinsky's team (major authors – Zarechnaya, Mikhaylushkin, Isaev, Mondal) have created interesting controversies, which we discuss below. A brief historical outline of works discussed here is given in Table 2.

**Table 2. Timeline and summary of work on γ-$B_{28}$ phase discussed in this review.**

| Reference | Main findings | Submission date | Publication date |
|---|---|---|---|
| Wentorf [9] | Synthesis, density measurement, qualitative X-ray diffraction, electrical conductivity change across β-γ transition. No chemical analysis and no structure | 04.10.1964 | 01.01.1965 |

| | determination. | | |
|---|---|---|---|
| Oganov et al. [7] | Synthesis and proof of chemical purity, structure determination, demonstration of partial ionicity, phase diagram. Introduced name γ-$B_{28}$ for the first time. | 27.01.2007 | 22.01.2009 |
| Solozhenko et al. [18] | Vickers hardness measurement (50 GPa). | 03.10.2008 | 01.12.2008 |
| Zarechnaya et al. [20] | Confirmation of synthesis and structure. Identified γ-$B_{28}$ with Wentorf's phase. | 03.11.2008 | 22.01.2009 |
| Le Godec et al. [25] | Measurement of the room-temperature equation of state up to 65 GPa. | 08.01.2009 | 22.05.2009 |
| Zarechnaya et al. [21] | Re-confirmation of structure using single crystals. Measurements of the band gap and electrical conductivity. Inaccurate measurements of the equation of state and hardness at 300 K. Incorrect interpretations of chemical bonding and density differences in boron phases. | 16.01.2009 | 08.05.2009 |
| Jiang et al. [26] | Simulation of the elastic constants and structure deformation mechanisms, supporting charge transfer picture. | 12.03.2009 | 11.05.2009 |
| Rulis et al. [27] | Simulation of electronic spectra, supporting charge transfer picture. | 06.04.2009 | 08.05.2009 |
| Zarechnaya et al. [272] | Claim of an isostructural transformation in γ-$B_{28}$, inconsistent with general theory of isostructural phase transformations and experimental evidence | 15.10.2010 | 18.11.2010 |
| Haussermann Mikhaylushkin [29] | Conclusion that there is no ionic contribution in bonding in γ-$B_{28}$, postulated unphysicality of Bader analysis. | 25.06.2010 | 20.12.2010 |
| Isaev et al. [30] | Experimental and theoretical confirmation of the Le Godec et al. equation of state [30]. Conclusion that DFT-GGA simulations accurately reproduces the equation of state. | 22.12.2010 | 21.04.2011 |
| Mondal et al. [31] | Confirmation, made using Bader analysis of an experimental charge density, of the results of Oganov et al. [7] on charge transfer in γ-$B_{28}$ | 29.11.2010 | 25.05.2011 |

## A. The equation of state (EOS), the proposed isostructural phase transition, and hardness.

Equation of state and "isostructural phase transformation". The EOS of γ-$B_{28}$ has been measured experimentally at 0-65 GPa [25], 0-30 GPa [21], 0-60 GPa [28] and 0-40 GPa [30]. Measurements and interpretations of Dubrovinsky's team show unusual features. It is already strange that two X-ray measurements of the zero-pressure density of γ-$B_{28}$ in two papers of Zarechnaya et al. [20,21] differ by 1%: 2.52 g/cm$^3$ [20] and 2.54 g/cm$^3$ [21], such an error being way above the quoted uncertainties of the measurement and probably coming from impurities. The samples were acknowledged to contain a mixture B+PtB [21], a result of a reaction between boron and platinum capsule; unfortunately, the concentration of Pt in the boron phase was not reported. Contamination is a serious concern, in view of the known extreme sensitivity of boron to impurities. High-pressure evolution of the density, i.e. the EOS, is even more controversial.

Theoretical calculations at the DFT-GGA level ($K_0$=241 GPa and $K_0'$=2.34 [30] at 300 K) agree well with some experiments ($K_0$=238 GPa [25,30] and $K_0'$=2.7 [25] or 2.5 [30]) but disagree with other experiments ($K_0$=227 GPa and $K_0'$=2.2 [20]). Zarechnaya et al. [28] claimed an isostructural transformation in γ-$B_{28}$ at ~40 GPa; below which the phase is more compressible ($K_0$=227 GPa) and above which much less compressible ($K_0$=281 GPa) than in previous experiments (Fig. 6). The suggestion of an isostructural transformation is inconsistent with all *ab initio* calculations and experiments, there is also no physical mechanism that could be responsible for such a transition, and the evidence for it given in [28] is self-contradictory.

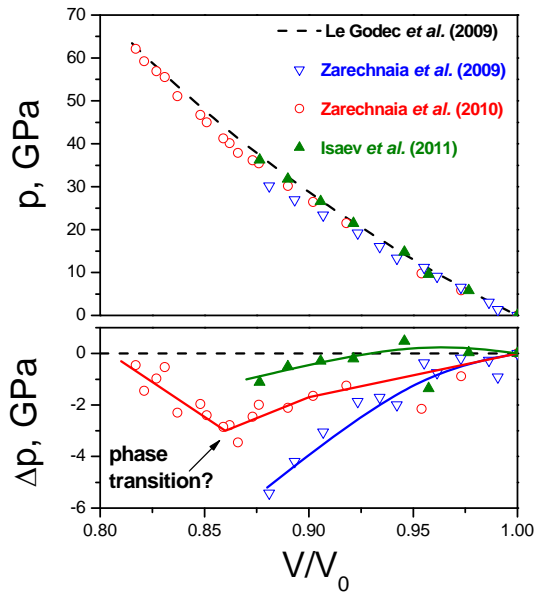

**Fig. 6.** Experimental equations of state of γ-$B_{28}$. (*Top*) Experimental data of different authors [20,25,28,30] (symbols) in comparison with the data by Le Godec *et al.* [30] (dashed line). (*Bottom*) Difference between the experimental equations of state [21,28,30] and that of Le Godec *et al.* [25] (dashed line). The latter data at 300 K can be described by a Vinet EOS with $V_0$=197.58 Å$^3$, $K_0$=238 GPa, $K_0'$=2.7. Static theoretical EOS of Oganov et al. [22] is described by a Vinet EOS with $V_0$=195.89 Å$^3$, $K_0$=222 GPa, $K_0'$=3.74.

Isostructural phase transitions (a subset of more general isosymmetric transformations, occurring with no change of space group) are well known in many systems and, while preserving symmetry and structure type, involve discontinuous changes in the density and/or electronic structure, which implies discontinuities in all vibrational frequencies – while Zarechnaya et al. [28] observed discontinuities only of some Raman modes (but not in the density or electronic structure), and continuous behavior of the rest. Only continuous behavior of Raman frequencies was reported in their own *ab initio* calculations.

As has been shown [32] using Landau theory, in agreement with experimental evidence, isostructural phase transitions must be first-order below the critical temperature and fully continuous above it. This involves a soft symmetry-preserving $A_g$ mode at the observed transition pressure. No such soft mode and no isostructural phase transition have been seen by any calculations done on γ-$B_{28}$ to date. Recent calculations of Isaev et al. [30] explicitly show the absence of any mode softening (instead, showing mode hardening) at 40 GPa both at 0 K and 300 K. Zarechnaya et al. [28] argued that the isostructural transformation is related to a kink in the pressure dependence of the LO-TO splitting parameter $\xi$:

$$\xi = \prod_{i=1}^{n} \left(\frac{\omega_i^{LO}}{\omega_i^{TO}}\right)^2 = \frac{\varepsilon_0}{\varepsilon_\infty} \quad , \tag{1}$$

which according to them characterizes the degree of polarity (i.e. ionicity) of bonding. In reality, the relationship of bond polarity and $\xi$ in crystals is not direct - $\xi$ originates from dynamical (rather than static) charges. If bond ionicity is negligible (as claimed by Zarechnaya et al. earlier [21]) and only dynamical charges are significant, the $\xi(P)$ dependence can affect only the phonon part of the equation of state, as can be easily shown, shifting it by not more than:

$$\Delta P_{max} = \frac{K_T E_{vib}}{6NV} \frac{d \ln \xi}{dP} \quad , \tag{2}$$

where $K_T$, $E_{vib}$, $V$ and $N$ are the isothermal bulk modulus, vibrational energy, volume, and number of atoms in the unit cell, respectively. Substituting here the values for γ-$B_{28}$, one obtains negligibly small $\Delta P_{max} < 0.02$ GPa, two orders of magnitude smaller than a typical uncertainty of pressure measurements. Thus, the explanation of Zarechnaya et al. [28] is unfeasible. In absence of a physically reasonable mechanism of the "isostructural transformation" and in the face of self-contradictory experimental data [28], one is forced to consider this transformation as an experimental artifact. What emerges from the majority of evidence, including the most recent theoretical-experimental work of Isaev et al. [30], is that DFT-GGA is capable of very accurately describing the EOS of γ-$B_{28}$ and other boron allotropes, and as theory suggests, that there is no isostructural phase transition in γ-$B_{28}$ within its stability field.

As an additional complication, Zarechnaya et al. [21] gave parameters ($V_O$, $K_0$, $K_0'$) of the equation of state without specifying the analytical form of the EOS (e.g. Vinet, Murnaghan, Birch-Murnaghan, etc), which renders their parameters practically meaningless. Fig. 7 shows that using the parameters [21] with three popular analytical forms of the EOS one gets very different results. For instance pressures are uncertain by as much as 10 GPa (Fig. 7a), and bulk moduli spectacularly diverge (Fig. 7b) at pressures where γ-$B_{28}$ is stable.

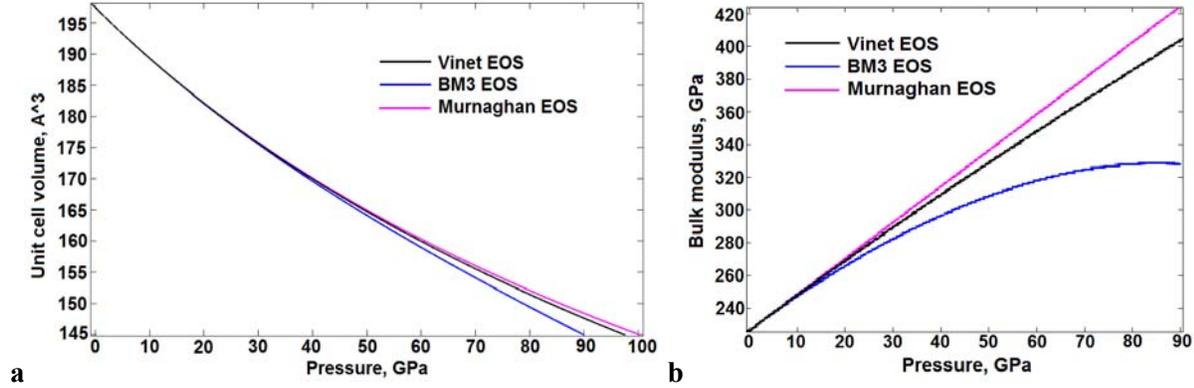

**Fig. 7. The importance of specifying the analytical form of the EOS:** (a) EOS and (b) bulk modulus as a function of pressure of γ-$B_{28}$, calculated using parameters reported in [21] and three different analytical forms of the EOS.

Hardness. The first value of Vickers hardness (50 GPa) by Solozhenko et al. [18] was followed by a substantially higher measurement of Zarechnaya (58 GPa) [21]. Incorrectness of the latter results transpires if one computes the hardness of γ-$B_{28}$ using various theoretical models: all results are more consistent with Solozhenko's value [18] than with Zarechnaya's result [21]. Existing theoretical models of hardness (see [33]) can be used to discriminate correct experimental results from artifacts. Jiang et al. [26] computed the equation of state, the elastic constants and remarkably high ideal tensile strengths (65, 51, and 52 GPa along the three crystallographic axes). The lowest ideal strength is often thought of as a good approximation to hardness – which in this case implies the hardness of 51 GPa. Subsequently, Zhou et al. [34] computed an even lower ideal strength for γ-$B_{28}$, for a discussion, see [35]. From the thermodynamic model of hardness [36] one gets 48.8 GPa. Using Chen's equation [37], the Vickers hardness is:

$$H = 2(\frac{\mu}{n^2})^{0.585} - 3 \,(\text{GPa}) \quad , \tag{3}$$

where $\mu$ is the shear modulus and $n$ is the Pugh ratio ($n=K/\mu$), and theoretical values of $K$=224 GPa and $\mu$=236 GPa [26], one obtains [37,38] the hardness of 49 GPa. Using the model of Li et al. [39] improved by Lyakhov and Oganov [40], for the main three boron allotropes, α-$B_{12}$, β-$B_{106}$, and γ-$B_{28}$, we obtain [40] the hardnesses of 39.9, 37.9, and 42.5 GPa, respectively - while the experimental values are 42, 45, and 50 GPa, respectively. To sum up, all theoretical models suggest the hardness of γ-$B_{28}$ to be in the range 49-51 GPa or lower, which is consistent with the experimental value of Solozhenko et al. (50 GPa - [18]), but not with that of Zarechnaya et al. (58 GPa - [21]).

## B. Is bonding in γ-$B_{28}$ partially ionic or fully covalent?

Ionicity and covalency: incompatible or ever-entangled? Covalent bonding originates from equal sharing of electrons (for instance, of an electron pair) between two atoms. When atoms are of different chemical types, electron sharing is unequal, with the bonding orbital asymmetrically shifted towards the more electronegative atom, leading to partial ionicity. The bonding orbital can be expressed in terms of atomic orbitals $\phi_A$ and $\phi_B$:

$$\Psi = c_A \phi_A + c_B \phi_B \quad , \tag{4}$$

where $c_A$ and $c_B$ are coefficients. This corresponds to the election density:

$$\rho = |\Psi|^2 = |c_A \phi_A + c_B \phi_B|^2 = c_A^2 \phi_A^2 + c_B^2 \phi_B^2 + 2 c_A c_B |\phi_A||\phi_B| \quad , \tag{5}$$

For a covalent bond, the coefficients $c_A$ and $c_B$ are equal, resulting in a symmetric charge distribution. Significant asymmetry and charge transfer appear when the two atoms have different properties. In Mulliken charge partitioning, the two atoms are assigned the number of electrons from the orbital (4) equal to $Z_A = c_A^2 + c_A c_B S_{AB}$ and $Z_B = c_B^2 + c_A c_B S_{AB}$ (where $S_{AB}$ is the overlap integral), respectively. If atom B has greater electronegativity than atom A, it also holds a greater part of the orbital $\Psi$ ($Z_B > Z_A$ and $c_B > c_A$), which can be represented as charge transfer from atom A (which now becomes a cation, a positively charged atom) to atom B (which now becomes an anion, a negatively charged atom). In this case we have a partially ionic bond; equivalent names for this situation are "polar covalent" or "mixed ionic-covalent" bond. The degree of ionicity can be estimated according to Pauling:

$$f = 1 - e^{\frac{-(\Delta\chi)^2}{4}} \quad , \tag{6}$$

where $\Delta\chi$ (= $\chi_A - \chi_B$) is the electronegativity difference. Both equation (6) and the expression for the molecular orbital (eqs. 4,5) imply that fully ionic bonding is not possible, and is a generally unattainable limit (although, for example, NaCl comes close to this limit). Neglect of this general and well-known fact sometimes leads to mistakes; for instance, recently, Zarechnaya et al. [21] and Mikhaylushkin and Haussermann [29] argued that γ-$B_{28}$ cannot be ionic because it is covalent. This not only incorrectly implies that covalent and ionic bonding are mutually exclusive, but also ignores the fact that any ionic bond has partial covalent character, and in most situations covalent bonds have a certain degree of ionicity (or, which is the same, polarity). Later [31] these authors essentially withdrew their earlier conclusions [21,29].

Equation (6) could be seen as implying no charge transfer between atoms of the same chemical type. However, one has to remember that equation (6) was proposed by Pauling as an interpolation over experimental data on dipole moments of diatomic molecules and has notable exceptions. First, it seems that charge transfer can be enhanced under pressure, as we have seen in Xe oxides [41]. Second, equation (6) not prohibits, but implies static charge transfer between atoms of the same type if these atoms have very different chemical environments (because electronegativity depends on the environment [42]). The latter happens in γ-$B_{28}$ [7].

Evidence for partial ionicity of γ-$B_{28}$. The structure of γ-$B_{28}$ may be seen, in a first-order approximation, as consisting of $B_2$ and $B_{12}$ clusters in a NaCl-type arrangement. Given that these two clusters have very different electronic properties, one can expect charge transfer between them. This was shown from several viewpoints, based on: (i) charge density analysis (in particular, Bader analysis), (ii) structural features, (iii) electronic structure (projected DOS, ELNES/XANES spectra), (iv) physical properties (in particular, lattice dynamics proves the existence of significant long-range electrostatic interactions between the atoms and LO-TO splitting).

There is no unique and universal definition of the charge of atoms in crystals, and therefore different definitions have been applied to γ-$B_{28}$ and yielded consistent results, indicating that this structure can be represented as

$(B_2)^{\delta+}(B_{12})^{\delta-}$ [7]. We obtained $\delta \sim +0.2$ from differences in the numbers of electrons within atom-centred spheres (sphere radii 0.7-1.0 Å); Born dynamical charges attain much higher values (spherically averaged $\delta=+2.2$). Our preferred estimate of charge transfer is based on Bader theory [43], which partitions the total electron density (ED) distribution into "atomic" regions separated by zero-flux (i.e. minimum-density) surfaces, and gives $\delta=+0.52$ (Table 1; the original result was slightly lower, $\delta=+0.48$). The exact values of the atomic charges depend somewhat on the level of theory - larger values of $\delta$ are obtained with hybrid functionals ($\delta=0.62$ with the HSE06 functional [44]) and with exact exchange ($\delta=0.68$). Among these, the most accurate values are probably given by the GGA [45] or HSE06 functionals. Bader partitioning is physically unbiased and ensures maximum additivity and transferability of atomic properties [43]. These relatively large Bader charges (certainly unusually large for a pure element) originate from the chemical interaction between the $B_2$ and $B_{12}$ clusters, as can be shown in two ways: the Independent Atom Model (IAM, where the total electron density is a sum of non-interacting spherically averaged atomic densities) has negligible charges (Table 1). Removing the $B_2$ pairs from the structure, we again obtain negligible charges (within ±0.03) even in the model interacting system composed of $B_{12}$ clusters in the same arrangement as in $\gamma$-$B_{28}$. Electron density on the partially ionic B1-B2 bond is very asymmetric, its bond asymmetry parameter reaches 20% [7]. Connected to this and arising from charge transfer, atomic volumes overall shrink for positively charged and expand for negatively charged atoms [7].

**Table 3. Bader charges (Q) and volumes (V) in $\gamma$-$B_{28}$ computed using the GGA functional, exact exchange and hybrid HSE06 functionals (this work), and two experimental determinations [31].**

|         | GGA   |         | Exact exchange |         | HSE06 |         | Experiments |         |
|---------|-------|---------|----------------|---------|-------|---------|-------------|---------|
|         | Q     | V, Å³   | Q              | V, Å³   | Q     | V, Å³   | Q           | Q       |
| B1 (4g) | +0.26 | 6.822   | +0.34          | 6.947   | +0.31 | 6.628   | +0.4138     | +0.8131 |
| B2 (8h) | -0.18 | 7.280   | -0.31          | 7.932   | -0.21 | 7.275   | -0.1943     | -0.1879 |
| B3 (8h) | +0.00 | 7.038   | -0.04          | 7.411   | +0.00 | 6.964   | +0.0571     | -0.0284 |
| B4 (4g) | +0.07 | 6.625   | +0.25          | 6.189   | +0.07 | 6.523   | -0.1386     | -0.4353 |
| B5 (4g) | +0.04 | 6.898   | +0.11          | 6.816   | +0.04 | 6.801   | +0.0011     | +0.0516 |

Structurally, $\gamma$-$B_{28}$ is similar to $\alpha$-$B_{12}$, but is denser due to the presence of additional $B_2$ pairs. The average intraicosahedral bond length is 1.80 Å and the B-B bond length within the $B_2$ pairs is 1.73 Å. The cationic $B_2^{4+}$ group is well known and its typical B-B distance (1.70-1.75 Å in $B_2F_4$ and $B_2Cl_4$) [46] is the same as in $\gamma$-$B_{28}$ (1.73 Å). On the other hand, the $B_{12}$ cluster is more stable as the $B_{12}^{2-}$ anion (as in the very stable icosahedral $(B_{12}H_{12})^{2-}$ cluster [46]), because in the neutral state it has an unoccupied bonding orbital (e.g., [47]). This orbital creates an acceptor band above the valence band edge in boron-rich solids. Electrons from dopant metal atoms or from other boron clusters may partially occupy this band, as detected by optical spectroscopy [48]. The $B_2$ pairs thus behave as electron donors, similar to the metal dopants in boron-rich borides. $\gamma$-$B_{28}$ is structurally related to several well-known compounds – for instance, $B_{12}P_2$ or $B_{13}C_2$, where the two sublattices are occupied by different chemical species (instead of interstitial $B_2$ pairs there are P atoms or C-B-C groups, respectively). This fact again highlights the chemical difference between the two constituent of clusters. This also gives one the right to call γ-$B_{28}$ a "boron boride" $(B_2)^{\delta+}(B_{12})^{\delta-}$ with partial charge transfer δ.

The computed atom-projected electronic densities of states show clear differences in local DOSs of the $B_2$ and $B_{12}$ clusters. These differences are even better seen when analyzing contributions of electrons with different energies to the total DOS (Fig. 8): the bottom of the valence band is dominated by $B_{12}$-icosahedra, whereas the top of the valence band and bottom of the conduction band are $B_2$-dominated. While this is consistent with significant charge transfer $B_2 \rightarrow B_{12}$, there is also a strong covalency in the system seen from strong hybridization in the middle of the valence band. In accord with this, Rulis et al. showed that the electronic spectra of the different atomic sites in γ-$B_{28}$ are indeed very different [27], in other words, different boron atoms on different sites behave as chemically different species, making charge transfer between them natural [7].

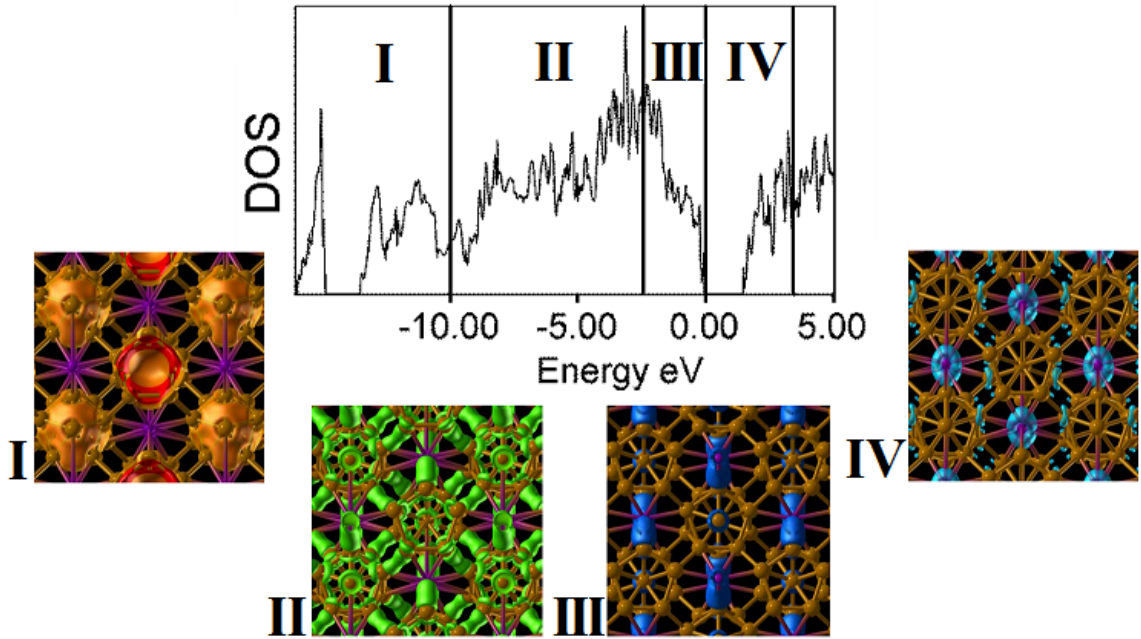

**Fig. 8. Electronic structure of γ-$B_{28}$.** The total density of states is shown, together with the electron density corresponding to four different energy regions denoted as I, II, III, IV. Note that lowest-energy electrons are preferentially localized around the $B_{12}$ icosahedra, whereas highest-energy electrons (including the bottom of the conduction band – "holes") are concentrated near the $B_2$ pairs. This is consistent with the direction of charge transfer: $B_2 \rightarrow B_{12}$. Modified from [8].

Despite considerable structural relationship with α-$B_{12}$ (*cf.* Figs. 1a and 1d), the electronic structure of partially ionic γ-$B_{28}$ is quite different: it shows little pressure dependence of the band gap and even at 200 GPa remains an insulator with a relatively wide gap (1.25 eV – and one has to bear in mind that DFT band gaps are usually seriously underestimated), whereas for the covalent α-$B_{12}$ the calculated band gap rapidly decreases on compression and closes at ~160 GPa (Fig. 9). The key for this different behaviour is charge transfer.

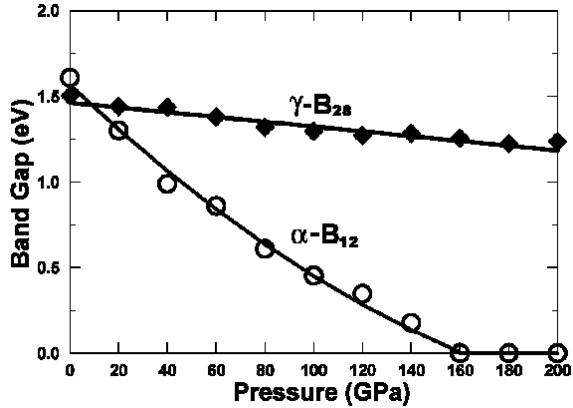

**Fig. 9. Pressure dependence of the band gap of α-$B_{12}$ and γ-$B_{28}$.** From [7].

Charge transfer (i.e. partial ionicity) affects many physical properties of γ-$B_{28}$, some of which are inexplicable within a purely covalent model. E.g., LO-TO splitting stems from long-range electrostatic interactions between the atoms: "the LO and TO modes … are *nondegenerate* for *ionic* crystals…, whereas they are *degenerate* for *non-ionic (*homopolar) crystals" [49]. The simplest parameter characterizing the LO-TO splitting is the dimensionless parameter $\zeta$ (eq. 1). $\xi$ = 1 for non-ionic crystals (the computed value is 1.01 for α-$B_{12}$) and $\zeta$ > 1 whenever there is charge transfer (1.16 for γ-$B_{28}$ and 1.18 for GaAs). Also, the computed high-frequency ($\varepsilon_\infty$) and static ($\varepsilon_0$) dielectric constants are very different (11.4 and 13.2, respectively), related to the LO-TO splitting and strong infrared absorption (which was indeed observed [7], in agreement with theory). Another (though less direct) evidence comes from the deformation mechanism predicted by Jiang et al. [26] – these researchers found that the first bonds to break are those between the most charged atomic positions, i.e. B1-B2 [26].

Swinging from one conclusion to the opposite, Dubrovinsky's team has created prolific, but controversial, literature on boron [20,21,28-31]. First, without estimating charge transfer, they claimed that it is zero in γ-$B_{28}$ [21]. Finding an accumulation of charge density between B atoms, they took it as a sign of fully covalent bonding without any ionicity – although as we discussed above, large covalent component of bonding does not imply the absence of ionicity, as mixed ionic-covalent (i.e. "polar covalent") bonding is extremely common and charge density accumulations are well known in such mixed ionic-covalent bonds as Si-O, which have a high degree of ionicity [50]. Then, Haussermann and Mikhaylushkin [29] (Mikhaylushkin was also a coauthor of [21]) concluded that there is no ionicity in γ-$B_{28}$ because, in their opinion (not supported by anything, but simply postulated) Bader analysis (used in [7] as one of the main tools) is meaningless. Just a few months later, a paper with Mikhaylushkin and Dubrovinsky as coauthors came out [31], which is based entirely on Bader analysis, finds large Bader charges with experimental electron densities (even larger than the theoretical charges found in [7]) and attributes absolute quantitative physical meaning to Bader charges. "Cognitive dissonance" contributes to multiplying controversies.

Rulis et al. [27] found that γ-B$_{28}$ does not satisfy the electron counting rules (more specifically, the Wade-Jemmis rule), which are fulfilled, for example, in the much more complicated β-B$_{106}$ structure. These authors concluded that the missing balance is compensated by charge transfer, i.e. partial ionicity. This conclusion was vehemently opposed by Haussermann and Mikhaylushkin [29], who concluded that electron counting rules can be satisfied if one considers the long B-B distances at 2.101 Å as bonds. A few months later, Mondal, Mikhaylushkin and coauthors [31], based on Bader analysis, showed (as was found earlier [7]) that the contacts at 2.101 Å should not be considered. They postulated that to satisfy electron counting rules, B1 atoms should attain a charge of +0.33, similar to Bader charges obtained in [7] and [31] (although the quality of Bader charges from [31] is hard to assess since their error bars were not given and two experimental estimates by the same authors gave a factor ~2 different results, see Table 3). While Ref. [29] described bonding in γ-B$_{28}$ as completely identical to that in α-B$_{12}$, the main point of Ref. [31] was that γ-B$_{28}$ has unique bonding with charge transfer between atoms of the same type, confirming the original results [7]. The main assumption and conclusion of the paper by Haussermann and Mikhaylushkin [29], was that if γ-B$_{28}$ satisfies electron counting rules, then it must be fully covalent – but this approach goes against basics of chemistry. The meaning of electron counting rules is in maximal/minimal filling of the bonding/antibonding orbitals and formation of a large HOMO-LUMO gap. As such, electron counting rules say nothing on whether the electrons are partitioned between the atoms equally (as in the fully covalent case) or unequally (as in cases with partial ionic character). For instance, as one increases the electronegativity differences in the series C→BN→BeO→LiF, the electron counting rules (the octet rule) remain fulfilled from purely covalent C (diamond, graphite, etc.), to mixed covalent-ionic BN, to highly ionic BeO and LiF.

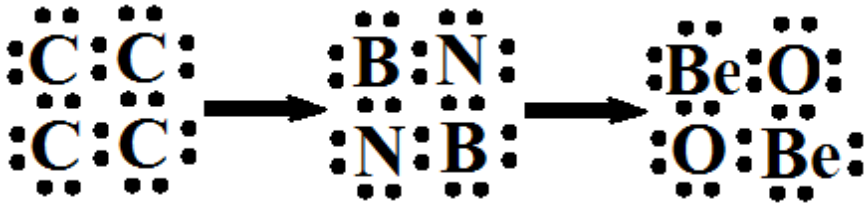

The bond valence sum rule of Brown [51] is another, more local, electron counting rule. Expressing bond valence as:

$$v_{ij} = \exp(\frac{R_0 - R}{\rho}) \quad , \tag{6}$$

where $R_0$ is a bond-specific parameter (1.5112 Å for B-B bonds) and $\rho$=0.37 Å. Brown (see [51] and references therein) demonstrated that in nearly all non-metallic crystal structures the sum of bond valences Γ on each atom equals its valence. While this rule is well fulfilled in α-B$_{12}$ (the B1 and B2 sites have Γ=3.04 and 2.96, respectively – very close to the valence of 3), γ-B$_{28}$ shows a different behavior (Table 4). For the icosahedral sites B2-B5 Γ=2.88-3.13, but the B1 site (even including the "fake" 2.1 Å bonds) is heavily underbonded, with bond valence sum Γ=2.35. This shows that local electron counting is actually not fulfilled, and that B1 and B2-B5 sites are chemically very different – contrary to [29].

**Table 4. Coordination and bond valence sums for boron sites in γ-B$_{28}$.**

| Site | Coordination number | Bond valence sum |
|------|--------------------|--------------------|
| B1   | 4                  | 2.35               |
| B2   | 7                  | 3.09               |
| B3   | 6                  | 2.88               |
| B4   | 6                  | 3.08               |
| B5   | 6                  | 3.13               |

Violations of the electron counting rules may indicate charge transfer (as suggested in [27]), but the opposite argument [29] is not physically correct. For compounds where electron counting rules are fulfilled, presence or absence of ionicity can be judged from other criteria – such as Bader analysis, LO-TO splittings, dynamical charges, dielectric constants, and detailed structural analysis, just as done in [7].

Summarizing the existing evidence, $\gamma$-$B_{28}$ possesses a unique type of chemical bonding, with a significant degree of ionicity in a pure element. This has eventually been recognized by initial opponents of this idea [31]. However, the idea itself can be traced to earlier literature – for example, similar suggestions existed for a high-pressure phase of hydrogen (proposed to contain $H^+H^-$ molecules [52]). For boron, the key is the ability of boron to form clusters with very different electronic properties.

To elaborate on this point, we have generated a number of hypothetical structures using the USPEX code [10,11], and among selected three structures shown in Fig. 10 and illustrating different aspects of this phenomenon. In the $P4/mmm$ structure (Fig. 10a), there are chains of octahedral clusters and single boron atoms. Atoms within the octahedral clusters have coordination numbers 5 and 8, whereas single atoms between them are 4-coordinate and have a highly negative Bader charge of -0.73. As expected, large differences in local geometry lead to large charge redistributions. The $Pm3$ structure (Fig. 10b) consists of $B_{12}$ icosahedra and single boron atoms between them. Just like in $\gamma$-$B_{28}$, the icosahedra carry an overall negative charge (-0.39), unequally distributed between atoms in the icosahedron (some are even positively charged) and single atoms have a large positive charge of +0.39. The $P6/mmm$ structure (Fig. 10c), like $\gamma$-$B_{28}$, contains $B_2$-pairs – which are combined here with graphene-like sheets of boron atoms. B-graphene sheets need one additional electron per boron atom to fill all bonding electronic levels. In this structure, $B_2$-pairs lose electrons and attain a charge of +0.58 (quantitatively similar to the charge of $B_2$-pairs in $\gamma$-$B_{28}$) and B-graphene layers develop an overall negative charge. The amount of charge transfer, although considerable, is less than what is needed to fill all bonding crystal orbitals, and the structure turns out to be metallic (just like low- and high-pressure phases of $MgB_2$ – see, e.g. [53]). In fact, all of the structures shown in Fig. 10 are metallic. They develop charge transfer because of large differences in the electronic properties of different boron substructures (or, in different words, charge transfer occurs to satisfy local electron count), but, because of very efficient screening in metals, do not have any long-range electrostatic interactions (leading, for example, to the LO-TO splitting).

$\gamma$-$B_{28}$, on the contrary, is an insulator, and displays both significant Bader charges and long-range electrostatic interactions between the atoms. Observations of dielectric dispersion [54], equivalent to LO-TO splitting, suggest similar phenomena in $\beta$-$B_{106}$, but detailed microscopic understanding for that phase is not yet available as reliable structural models of β-B$_{106}$ are only now beginning to emerge from computational studies [4-6]. In a pure element ionicity can appear only as a result of many-body interactions, which distinguishes it from classical ionicity.

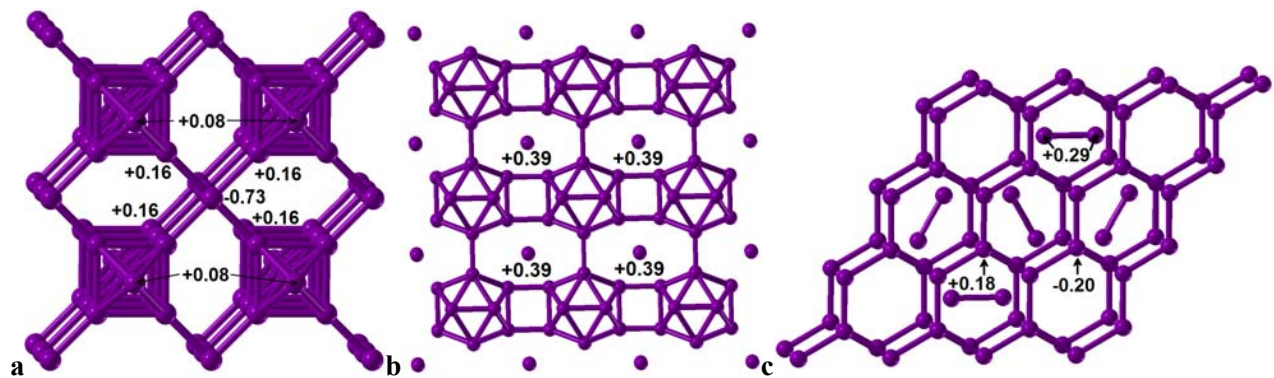

**Fig. 10. Hypothetical structures of boron showing large Bader charges:** (a) *P4/mmm* structure consisting of chains of corner-sharing B$_6$ octahedra (atomic charges +0.08 to +0.16) and single atoms bonded to them (atomic charge -0.73), (b) *Pm*3 structure composed of B$_{12}$ icosahedra (atomic charges from -0.08 to 0.02) and single boron atoms (atomic charge +0.39), (c) *P6/mmm* structure that can be represented as an AlB$_2$–type structure where 3/4 of Al positions are occupied by B$_2$ pairs. In other words, in this structure B$_2$ pairs (with each atom carrying the charge of +0.29) are sandwiched between B-graphene sheets (in which the atoms have charges of -0.20 and +0.18). The values of Bader charges are shown.

**III. Conclusions and outlook.**

Largely due to complicated chemistry, experimental studies of boron proved to be highly non-trivial, often leading to erroneous results even with modern methods (see [22]). The history of studies of boron has many additional examples of this [8]. Major progress has been achieved with the discovery of γ-B$_{28}$, [7,18], but one should never forget that boron is still an element of surprise, and many aspects of its behavior remain enigmatic. Out of 16 allotropes that have been reported in the past, 4 (including γ-B$_{28}$) are now confirmed to be thermodynamically stable pure boron allotropes. It is hard to imagine that these four phases encompass the entire structural variability of this element, and we expect discoveries of new allotropes with new twists of chemistry and interesting physical properties. γ-B$_{28}$, with its relatively simple and beautiful structure, as well as attractive properties and unique chemical bonding, has attracted tremendous attention in the literature. New chemical thinking that is being produced by such studies, and by this tractable material (compared to incomprehensibly complex disordered β-B$_{106}$ and T-192 structures) is likely to lead to future breakthroughs in chemistry and materials science. Other reviews in this Special Issue [35, 55-59] consider various aspects of physics and chemistry of γ-B$_{28}$ and related compounds. It is indeed amazing to see how much work has been done on this material since 2006!

This work is supported by DARPA (grant N66001-10-1-4037). We thank P. Macchi for useful discussions.